# Near-Optimal Virtual Machine Packing Based on Resource Requirement of Service Demands Using Pattern Clustering

Yaghoob Siahmargooei*, Mohammad Kazem Akbari,
Seyyed Alireza Hashemi Golpayegani
Department of Computer Engineering and IT
Amirkabir University of Technology (Tehran Polytechnic)
Tehran, Iran
{y.s, akbarif, sa.hashemi}@aut.ac.ir

Saeed Sharifian
The Electrical and Electronics Engineering Department
Amirkabir University of Technology (Tehran Polytechnic)
Tehran, Iran
sharifian_s@aut.ac.ir

*Abstract*—**Upon the expansion of Cloud Computing and the positive outlook of organizations with regard to the movements towards using cloud computing and their expanding utilization of such valuable processing method, as well as the solutions provided by the cloud infrastructure providers with regard to the reduction of the costs of processing resources, the problem of organizing resources in a cloud environment gained a high importance. One of the major preoccupations of the minds of cloud infrastructure clients is their lack of knowledge on the quantity of their required processing resources in different periods of time. The managers and technicians are trying to make the most use of scalability and the flexibility of the resources in cloud computing. The main challenge is with calculating the amount of the required processing resources per moment with regard to the quantity of incoming requests of the service. Through deduction of the accurate amount of these items, one can have an accurate estimation of the requests per moment. This paper aims at introducing a model for automatic scaling of the cloud resources that would reduce the cost of renting the resources for the clients of cloud infrastructure. Thus, first we start with a thorough explanation of the proposal and the major components of the model. Then through calculating the incomings of the model through clustering and introducing the way that each of these components work in different phases, we mention the way that the required outcome which is the decision made for the number of proper virtual machines and the appropriate configuration and applying of services on virtual machines. Later, by testing through the produced data, we apply the model. Finally, we will analyze the results of testing and evaluating the model. The final quantity will be evaluated appropriately so that we could prove the accuracy of the application of the model.**

## I. Introduction

Cloud can be defined as a large-scale distributed computing paradigm where a pool of virtualized, scalable, and manageable computing power, storage, platforms and services can be provisioned on-demand to customers over the Internet [1].

Compared with traditional desktop computing, cloud computing presents many advantages, such as better resource utilization, rapid elasticity, higher power conservation and economies of scale, which can save the up-front investment of enterprise information system and reduce the daily operation and maintenance costs significantly in the long run [2]. With the advancement of cloud computing technologies, including virtualization, security, Service-Oriented Architectures and high bandwidth network access, it is becoming a trend that large numbers of existing business applications from companies and institutes will be migrated into clouds and deployed as cloud services due to the above-mentioned benefits [3,4]. In Cloud computing, customers can avoid capital expenditure on hardware and software by renting the usage from the service provider of a third party, rather than owning the physical infrastructure by themselves. The hardware and software are rendered to customers as IT services [5].

Consider an E-Commerce Organization which has rented cloud computing services in order to meet its processing needs. With regard to its organizational processes and workflow, this organization needs different services for presentation of tools and electronic services. these services are placed in the workflow as job duties. Furthermore, this organization pays money for the resources that it has rented from cloud Computing service providers. Now the question is that how much of these processing resources and cloud infrastructures are needed in order for meeting the processing needs of this organization. Another question is: what are the ways through which the organization can approximate its current processing resources to its required processing resources so that it can reduce the cost of the rents in a way that it would not lead to such problems as the lack of resources and lack of the customers' satisfaction with the services and high response time for providing services? Moreover, the service requests of the organization might undergo radical changes which make the situation far more difficult. For instance, in Fig. 1 you can see the diagram of the number of the users' visits of a Football World Cup website in different hours of the day in 1998. Such a workload is very typical of all commercial websites and planning capacity for such workload is not easy.

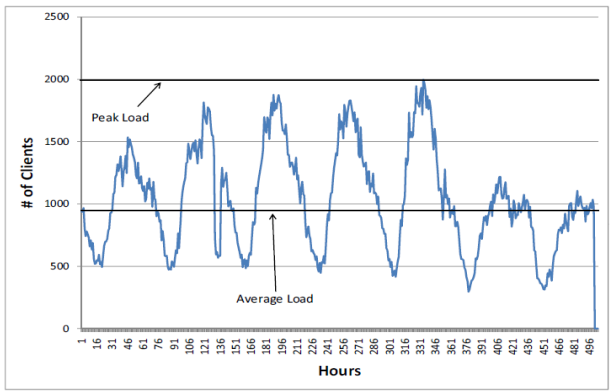

Figure 1.  Incoming traffic of the website of 1998 Soccer World Cup

Capacity could be planned for the average load as shown in Fig. 1 or for the peak load. Each approach has its disadvantages, however [6]. In order to maintain the quality of service, the managers of the system must provide enough resources to confront with the request changes in workflow. Unfortunately, repeated provisioning of the resources would heighten the costs and would thus lower the profit in business. In other words, lack of providence would diminish the quality of service, irritation of the clients and consequently would diminish their persistence in business. Traditionally, providing the hardware in monthly or annual periods were done based on the politics of the company and the expected workflow. As you see in Fig. 2, most of the time an upgrading of the hardware as a result of an increase in providence (and in other words, loss of money) happens at a time when there was a plan for running a new hardware was during the next months or years. Of course in some periods of time, the present hardware cannot manage to meet the real demand, which would diminish the quality of service [7]. Lack of provisioning of the resources would diminish the performance and extreme provisioning, on the other hand, would waste the resources and heighten unnecessary costs [8].

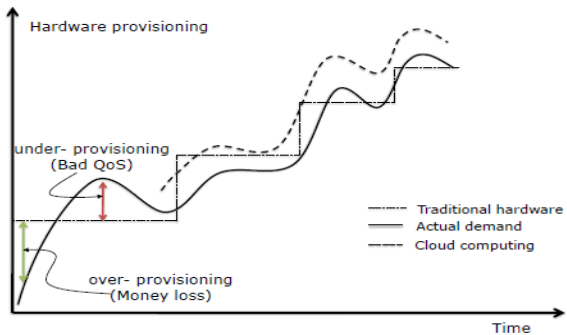

Figure 2.  Diagram of the process of upgrading of the hardware and its providence throughout time [3]

The most recent provisioning technology in virtual machines is capable of providing a virtual machine in 10 minutes. Such delay, for those works which require scaling in the process of calculation, is not acceptable; for the purpose of scaling at the moment, the virtual machines should get ready in a few seconds. Of course from the perspective of hardware and software technology this is an impossible dream that we could decrease the duration of virtual machine providence. From business perspective, the simple way is to ask the clients their calculating needs for scheduling the resources but the problem is that the clients do not know their calculating needs and their resources and could not provide a proper schedule; besides the clients' needs are always changing that is why we cannot come to an accurate evaluation of the needed resources and its schedule. Having technological and business problems in mind, the only solution to this problem is the anticipation and provisioning of virtual machines beforehand [9].

Cloud provider proposes different kinds of virtual machines with different powers for different applications. Based on what is there in the real world, the cost of setup of virtual machines is calculated per hours. Besides, the virtual machines could be provided at any time; there is just a little time needed so that a virtual machine will be setup. We have considered this amount of time as estimative and we believe that if we could get access to the patterns in desired time, we could finish the job within a certain amount of time [4]. Most of the cloud providers design two payment methods, (1) Reserve or prepaid and (2) Payment based on demands. The reserve method is usually cheaper than payment based on demands, but in the case of reserve there is need for provisioning for the purpose of development. The reserve method is very useful if the real calculating demands are specified though in reality the real calculating demands is noticeable only when it is in use. Therefore, the accurate reservation of the amount of resources is very hard at the beginning [10].

## II. RELATED WORK

Currently, cloud providers propose methods based on scheduling and methods which are rule-based for the purpose of scaling up or down. Many of these methods are formed based on the application of resources like the rule: "if average utilization of the processor is and more than 70% for more than 5 minutes, you should run two new instances." this method is simple and appropriate, though in many cases it is not possible for the users to specify the utmost and threshold of scaling especially when the program has a complex model and the amount of resources is very low. In other words, this method is not performable enough for specifying scaling [8].

If we want to have a complete categorization of the completed works in this area and related to this topic, we can categorize the suggested methods into five categories that you will see [11]:

1) *Static model and the methods based on threshold.*
2) *Reinforcement learning.*
3) *Queuing theory.*
4) *Control theory.*
5) *Analysis of periods of time.*

If we want to have a short definition of these categories, we can say that in static method and threshold based, Upon the change of rules and conditions which work based on specific conditions and thresholds, we can change the state of the system and status of the resources. Reinforcement learning method, which is a king of automatic decision-making, is through reviewing the status of performance of the cloud resources and without having special priorities in an attempt to learn and solve the problems. Queuing theory is through using classic mathematics problem, that is, Queue attempting to metric estimating of the performance of the resources such as the length of Queue and the duration of interval of requests and using different lineups for a better solving of the problem. In control theory methods it is attempted to solve the problems through using prevention or active methods through different controlling done on major components of the cloud resources. The last category, which is used in different areas such as economics, engineering and bioinformatics, tries to suggest some solutions through using the applied changes in different periods of time [11]. The important points in auto-scaling of the resources are: (1) Changing the number of resources, (2) Ability to anticipate workflow, (3) And calculating the amount of required resources at the time when workflow increases or decreases. Furthermore, another important area in auto-scaling is specifying of the resources by considering the optimizing of different factors of cost.

### III. PROPOSED MODEL

In order to solve this problem and auto-scaling of the cloud resources, one can deduce the number of incoming request in every service, through considering the amount of changing requests in different periods of time and through monitoring incoming requests to the system. This is done in a way that by putting these quantities together, we would come up with a request pattern like Fig. 3. This pattern shows how many requests we have in every t period of time.

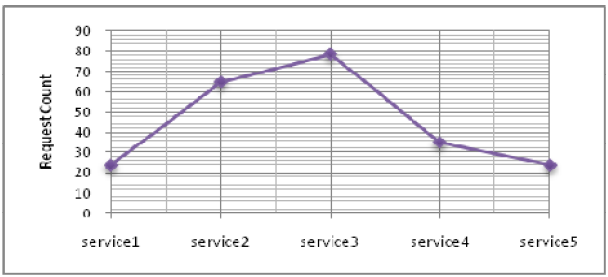

Figure 3. Method of calculating request pattern based on service requests

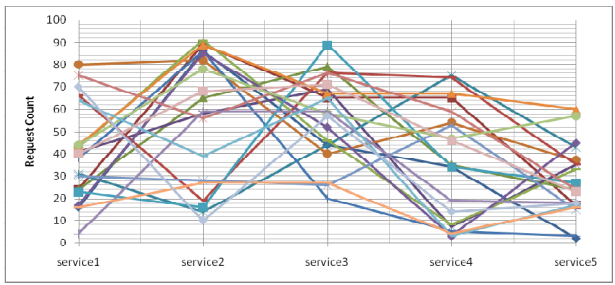

Figure 4. Displaying periodical request patterns at a whole

Now by putting these patterns together in a specific period of time like a day or a month, we can have a collection of incoming requests patterns which we could be seen in Fig. 4. Considering the fact that each of these patterns are made based on the behavior of the applicants and more accurately through using modeling graph of the customers' behavior, one can find a close similarity between them. The reason is that some of the patterns are very much similar to the others in terms of the number of service requests and one can put them in specific categories. Now, one can find a specific number of these categories which have almost similar patterns through clustering. If we can calculate a specific pattern through the quantity of the resources, and if we could accurately estimate the number of needed virtual machines, we can introduce a new method through a table which is composed of the centers of the clusters of the patterns and the number of its needed virtual machines.

Before analyzing the phases of the work, we need to accurately analyze the mentioned points. Through considering periods of time as t and the number of incoming requests of each of these services in every single period of time as Ds,t and considering the amount of used resources in each service for responding to a request as Ns,t, one can calculate the amount of needed resources of service S in a t period of time. This amount which is calculable through (1) is called Rs,t. through putting these quantities together, a vector is made for all of these services which is called vector of request pattern in a period of time. This vector is expressed through (2):

$$R_{s,t} = \sum_{i=1}^{S}(D_{i,t} \times N_{i,t}) \quad (1)$$

$$Demand\ Vector_t = [R_{s1,t} \quad R_{s2,t} \quad ... \quad R_{sN,t}] \quad (2)$$

Now in a specific period of time we have plenty of request pattern vectors which, as it was mentioned before, could calculate the present clusters through common clustering methods such as SOM or AHC. It should be mentioned that, in case the number of the needed clusters are definite from the very beginning, one can use K-Means algorithm for clustering. Through calculating the centers of the clusters, we come to a final number of request patterns which we call as representative request patterns which are the request patterns with the highest repetition. As you notice in Fig. 5, representative request patterns show a high repetition of a special status of the system or a special request pattern by the customers and users. Now we need to calculate how many virtual machines and processing resources and what kind of configuration of services we need on each of them. This, which should be done with a high accuracy and offline and meta-heuristic algorithms, need a lot of searching in solution space so that they could find a solution which is nearer to optimum in a vaster space.

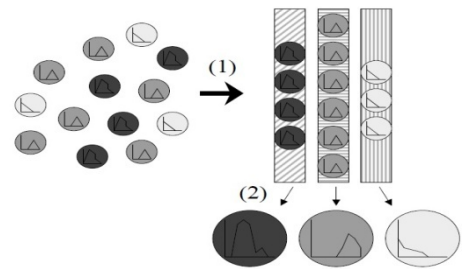

Figure 5. Method of clustering periodical request patterns using K-Means

One of the advantages of the proposed method is lack of need to calculate on line. The reason is that the mentioned methods of calculation are applicable in offline mode and also one can calculate the required processing resources and optimum configuration in specific periods of time through finding representative patterns in a reasonable time. Doing this, we can make a look-up table for all representative patterns which are considered to be the centers of our clusters. This table, as you can see in Fig. 6, consists of two columns in the first of which there are representative patterns and in the second of which there are parallel configuration and the required processing resources. Since we need a constant need to look up this table, we call it look-up table. Pay attention that the configuration of services on virtual machines are done for the purpose of load balancing and based on the number of received demands from those services and the amount of required processing resources; and with regard to its performance or lack of its performance in a specific virtual machine the quantity of binary would be 0 or 1. Virtual machines would be identified through a single number and the characteristics of their resources.

| Pattern | Assignment |
|---|---|
| [25,60,12,32,48] | [VM1,1,1,2][0,1,1,0,0] |
|  | [VM2,1,2,1][1,1,1,0,0] |
|  | [VM3,2,1,2][1,0,0,1,1] |
| … | … |
| [10,12,17,16,13] | [VM1,1,1,2][1,1,1,1,1] |
|  | [VM2,1,2,1][0,0,1,1,0] |

Figure 6. Lookup Table

without paying attention to specific periods of time in online model, now we can deduce processing resources and the number of required virtual machines in the current situation, only through calculating the vector of incoming request pattern and through using a simple comparison of incoming request pattern with the current request patterns in the lookup table. For comparing the two patterns, Euclidean similarity measures and correlation of the two patterns are used. It should be mentioned that this not only would not need a long time but also would need only little calculation at the moment which is one of the innovations of this method. In order to finalize this section, the phases of application of the proposed model are applicable as follows:

*1)* Collecting the service requests in different periods of time as primary incoming data.

*2)* Calculating the quantity of required processing resources.

*3)* Calculating request patterns of the services through using the last step.

*4)* Clustering of all service request patterns.

*5)* Identifying the centers of cluster as representative patterns.

*6)* Calculating the required processing resources as representative patterns.

*7)* Making a lookup table based on the calculations of the previous steps.

*8)* Using the model in an online mode and comparing the incoming pattern.

*9)* Sending different request patterns for clustering (returning to step 4).

*10)* Announcing appropriate configuration of virtual machines to cloud provider.

## IV. TESTING THE MODEL

In order to apply the proposed model and performing the required tests in order to evaluate its performance, an E-Commerce website was considered which has a special workflow and uses 5 electronic services. Periodical efficiency of the model was considered to be 10 minutes. Now we deduced the incoming traffic of the website in one period of 100 time periods as its history. This was deduced through the report available at Tomcat server of the website; then the preprocessed data were used in order to produce service request patterns. Now having 100 service request patterns, we started clustering the patterns through using AHC and K-means algorithm. As you notice in Fig. 7, clustering was done for selecting 10 clusters the results of which and the shapes of the centers of the clusters of which could be noticed in Fig. 8. Considering 10 clusters were done based on applying goodness tests on the clusters with Davis-bouldin index and Dunn index. Thus, the best number of the clusters, that is 10 clusters, is justifiable.

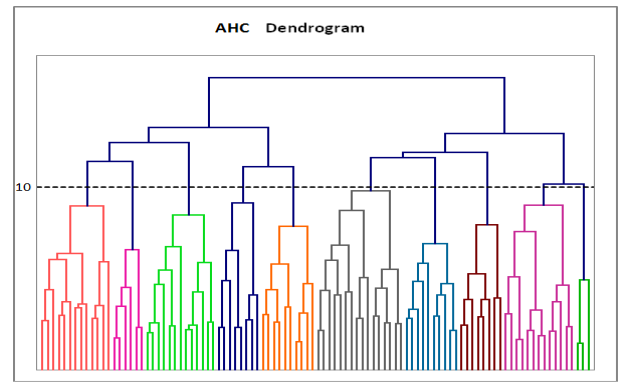

Figure 7. Dendrogram diagram clustering for 10 clusters

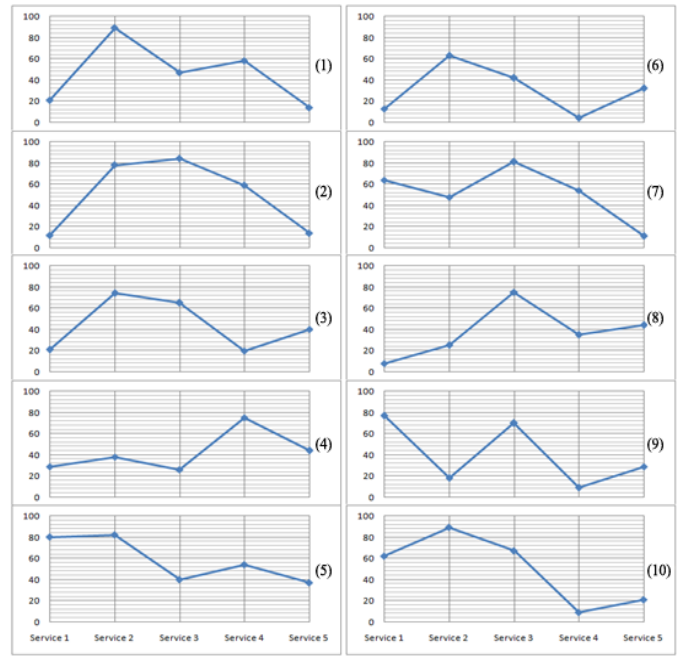

Figure 8. Diagram of the centers of the clusters after clustering

The centers of the clusters are representatives of the clusters belonging to the same cluster, which has an extreme similarity to the other members of the cluster. In Fig. 9, for instance, you can notice the center with the members of the first cluster. The difference between these members is very minor and thus one can ignore the differences.

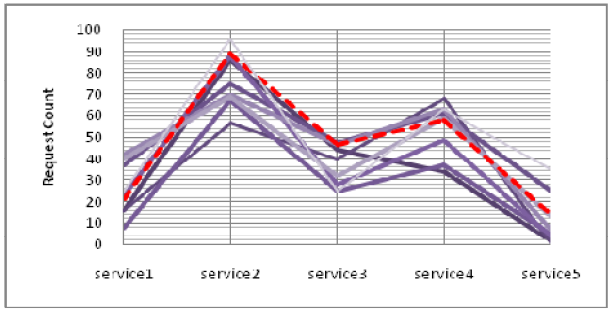

Figure 9.   Request patterns in a cluster

Further, as it was mentioned in the previous section, through using meta-heuristic algorithm like genetic algorithm for representative patterns resulted from clustering, the optimum number of virtual machines were considered based on the quantity of required resources of that service and the processing power of that machine. After adjusting accuracy of the parameters of genetic algorithm through using Taguchi method and doing complicated calculations in order to find the best fitness function, we tried to make lookup and configuration table ready and then we took the system to an online mode ready for applying the proposed model. With the introduction of the new incoming traffic, request pattern parallel to that period of time would be calculated through (1). Then through using the quantity of correlation between the incoming pattern and the current patterns in lookup table, the best (in terms of similarity) representative pattern would be chosen. As you notice in Fig. 10, the most similarity between the incoming pattern and the current representative patterns present in the lookup table belongs to cluster number one. The quantity of correlation for this pattern is 0.91. In this diagram, the pattern with bold square nods is the representative pattern number one and the pattern with diamond nods is incoming request pattern in online mode. The similarity between the incoming patterns with the selected pattern is quite visible even superficially.

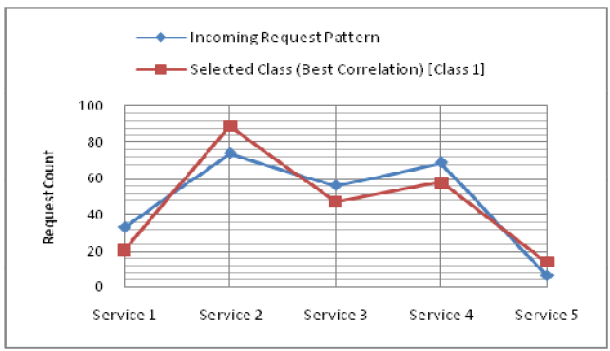

Figure 10. Diagram of choosing the best representative pattern for the incoming pattern

Now through going to the lookup table and taking the second column parallel to the incoming pattern, the number of required optimum virtual machines and the proper configuration of the services for the purpose of load balancing will be chosen and announced to the cloud provider. It should be mentioned that there are instances where the incoming pattern has no similarity with the current patterns in lookup table. The reason is that the representative patterns deduced from clustering of patterns of special periods of time and it is possible that in a specific situation, a different request would appear. In Fig. 11 you can notice an example of an incoming request pattern which has low similarity with the selected pattern after comparing the correlations. The quantity of correlation for this pattern is 0.56. considering the fact that the selected pattern contains the most similarity to the incoming pattern, it is important that with considering a threshold amount, those patterns which have a low similarity to the representative pattern will be collected and added to clustering and the centers of cluster in the next periods so that they will be added to lookup table with a proper configuration, in case it is necessary. This would not only increase the accuracy of the model but also would lead to its being up-to-date and its gradual improvement. That is why, the quantity of threshold for correlation quantities is considered to be 0.7. if the utmost quantity of correlation of request patterns with representative patterns will be less than 0.7, the incoming pattern for the next clustering will be saved.

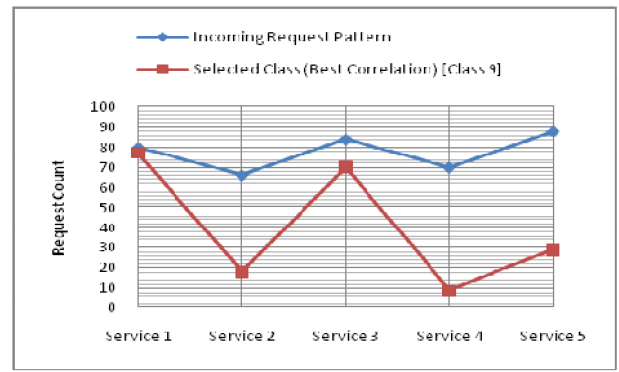

Figure 11. The quantity of low correlation between the incoming pattern and the selected representative

## V.  RESULTS AND CONCLUSION

As it was mentioned in the previous section, through clustering the incoming request patterns and calculating the number of optimum virtual machines and a proper configuration of the services for the centers of clusters and representative patterns, we have introduced a new method for auto-scaling. It should be mentioned that request patterns are different for different resources since each service from each resource needs a different quantity. Upon considering more processing resources, calculations and clustering would get more complex. Furthermore, if we use correlation measure for comparing the similarities between these patterns we need to consider that scale for virtual machines for those patterns which have a high correlation (almost 1) but are different in terms of number; and if it is possible we use another similarity measure such as Euclidean distance. Besides, if there is a special pattern which has the least similarity with the current patterns in a specific period of time and the present representative patterns in lookup table, the number of virtual machines and configuration would not be optimized which is one of the limitations of work.

In order for the betterment of the work and evaluation of the introduced model, we need to identify the new incoming patterns (with considering the cost of the rent of resources and cost of the final quantity of virtual machines in the lookup table) through offline method rather than through comparing with lookup table. Now we can compare the deduced quantities with the current quantities in the parallel lookup table; based on Fig. 12 because of considering the utmost quantities for the service requests of the resources and because of considering the worst case scenario for specifying the services, one can say that the results of evaluation show good performance of the model. Moreover, based on the results in Fig. 13 by comparing the outcome of offline algorithm which uses meta-heuristic methods, more difference and a better improvement would be noticed through greedy methods such as Best Fit and First Fit.

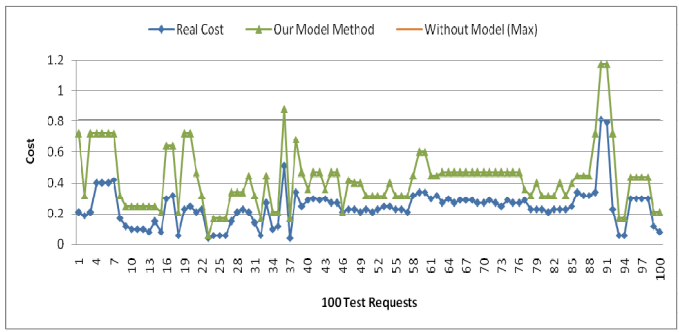

Figure 12. Comparing the calculated cost for 100 incoming requests with different methods.

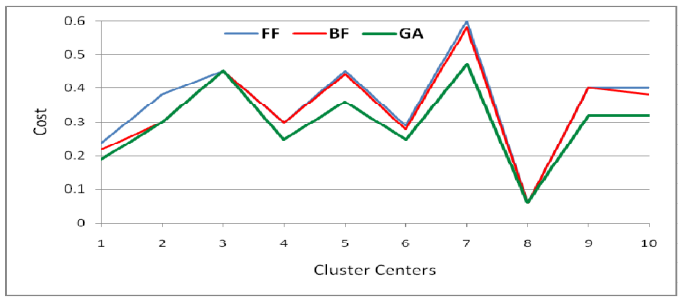

Figure 13. Comparing the calculated cost by the proposed algorithm through greedy algorithms.

In this paper through proposing a model for auto-scaling of virtual machines in cloud computing environment through clustering request patterns and considering workflow and organizational services, the number of required momentary virtual machines and also the cost of renting resources were decreased. Additionally, we focused on the method of deduction and making request pattern and their clustering in order to deduce representative request patters based on them through using focused tests and we briefly mentioned offline algorithm for calculating optimum numbers of virtual machines and the configuration of services on each machine through using meta-heuristic methods.